\documentclass[letter]{aa}
\usepackage[varg]{txfonts}
\usepackage{lineno} 
\usepackage{amssymb,amsmath}
\usepackage{graphicx}
\usepackage{xcolor,natbib}
\usepackage{multirow}
\usepackage[T1]{fontenc}
\usepackage{ae,aecompl}
\usepackage{newtxtext, newtxmath}
\usepackage{float}
\usepackage{subfigure}
\usepackage{longtable}
\usepackage{enumitem}
\usepackage{longtable,listings}
\usepackage[flushleft]{threeparttable}
\usepackage{parcolumns}
\usepackage{hyperref}
\leftlinenumbers               
\usepackage{CJK}
\usepackage{upgreek}

\begin{document}

\title{Time-dependent obscuration of a tidal disruption event candidate in the active galactic nucleus CSS100217}

\titlerunning{An obscured TDE in CSS100217}

\author{Ying Gu \inst{\ref{inst}}
\and Xiao Li\inst{\ref{inst}}
\and Xing-Qian Cheng\inst{\ref{inst}}
\and Dou-Dou Wang\inst{\ref{inst}}
\and Xue-Guang Zhang\inst{\ref{inst}}
\and En-Wei Liang\inst{\ref{inst}}}

\institute{Guangxi Key Laboratory for Relativistic Astrophysics, School of Physical Science and Technology, Guangxi University, Nanning 530004, China\label{inst} \\
\email{xgzhang@gxu.edu.cn; lew@gxu.edu.cn}}
\date{Received XXX / Accepted XXX}

\abstract
{CSS100217 is considered a peculiar tidal disruption event (TDE) candidate occurring in an active galactic nucleus (AGN). Unlike typical TDEs, for which the post-flare luminosity is equal to that pre-flare, CSS100217 decayed to $\sim$ 0.4 magnitudes fainter than its pre-flare V band level. In this manuscript, we propose an obscured TDE model to explain the light curve of CSS100217.
Assuming that the time-dependent obscuration, caused by the TDE unbound stellar debris, or by nuclear clouds moving around the supermassive black hole (SMBH), follows a Weibull distribution, we find that the light curve of CSS100217 can be described by the tidal disruption of a $4.6_{-0.9}^{+0.9}{\rm M_\odot}$ main-sequence star by a $3.3_{-0.3}^{+0.3}\times10^7{\rm M_\odot}$ black hole. The total energy of the event derived from our fit is $7.23\times10^{53}$ ergs and about 1.38 ${\rm M_\odot}$ of debris mass is accreted by the central SMBH. The model indicates that the contribution of the host galaxy to the observed pre-flare optical luminosity is not-significant compared to that of the AGN, which is consistent with the results of the spectral analysis. These results suggest that obscuration may play an important role in explaining the unusual TDE-like variability observed in CSS100217.}

\maketitle

\section{Introduction} \label{sec:intro}
A tidal disruption event (TDE) can be expected when a star wanders around a central massive black hole (BH) with a distance to the BH smaller than the tidal disruption radius ($R_t\sim R_*(M_{\rm BH}/M_*)^{1/3}$ with $R_*$ and $M_*$ the radius and mass of the star, respectively, and $M_{\rm BH}$ the central BH mass). Since the theoretical concept of TDEs was first proposed \citep{1975Natur.254..295H,1988Natur.333..523R}, an increasing number of theoretical simulations have been performed (e.g., \citealt{2020SSRv..216...63L,2023ApJ...946L..33R,2024Natur.625..463S}), and more and more TDE candidates have been detected and reported in the literature (see e.g., the samples presented in \citealt{2021MNRAS.508.3820S,2021ApJ...908....4V,2021ARA&A..59...21G,2023ApJ...955L...6Y,2025A&A...697A.159G}).
In the classical picture of TDEs, once a star is disrupted, about half of the tidal debris becomes gravitationally bound to the supermassive black hole (SMBH) and powers a luminous, multiwavelength flare through accretion.  The other half is unbound and escapes to infinity on hyperbolic orbits. The light curve of the TDE typically rises to peak over weeks to months, then decays following a power law of $t^{-5/3}$  (e.g., \citealt{1988Natur.333..523R,2019GReGr..51...30S}). However, deviations from this canonical behavior may occur due to a variety of physical processes (e.g., \citealt{2009MNRAS.392..332L,2013MNRAS.434..909H}). At later times, the decline flattens and eventually returns to the pre-flare level.

Among the reported TDE candidates, especially those identified in the optical, almost all have been detected in inactive galaxies (e.g., \citealt{2020ApJ...891...93F}).
This is mainly because identifying TDEs in active galactic nuclei (AGNs) is more difficult due to their inherent variability and flare-like features (e.g., \citealt{2013ApJ...767...25G,2017MNRAS.470.4112G,2023ApJ...942....9H}). So far, only a few TDE candidates have been detected in active galactic nuclei  {\citep[AGNs;][]{2015ApJ...800..144L,2021MNRAS.500L..57Z,2022MNRAS.516L..66Z,2023A&A...672A.167H}.
Additionally, an increasing number of large-area sky surveys have discovered unusual accretion events in AGNs that resemble TDEs. However, their origins and physical mechanisms cannot be definitively determined (e.g., \citealt{2017MNRAS.470.4112G,2019NatAs...3..242T}).

CSS100217:102913+404220 (hereafter CSS100217) is a superluminous transient that appears in the central regions of a narrow-line Seyfert 1 galaxy (NLS1) at a redshift of $z=$  0.147. CSS100217 was initially interpreted as a supernova \citep{2011ApJ...735..106D}, while further in-depth analysis of follow-up observations concluded that it is more consistent with a TDE or an AGN scenario \citep{2017ApJ...843L..19M,2017ApJ...843..106B,2017NatAs...1..865K,2021ApJ...920...56F,2022A&A...660A.119Z,2022MNRAS.516..529C}. Interestingly, the optical light curve of CSS100217 exhibits a long-term low-luminosity state coincident with the decay of the flare, with the post-flare brightness about 0.4 magnitudes fainter than the pre-flare level \citep{2017ApJ...843L..19M}.
\cite{2022MNRAS.516..529C} suggest that the behavior of the post-flare light curve of CSS100217 can be attributed to the tidal disruption of a star on a retrograde orbit with respect to the accretion disk rotation. The resulting stream of disrupted material perturbed the disk, creating an empty cavity that has yet to be replenished, thereby leading to the observed low-luminosity state.

In this paper, we focus on the post-flare low-luminosity state of CSS100217. We propose for the first time that the atypical TDE light curve behavior of CSS100217 may result from a scenario in which the unbound stellar debris from the TDE, or moving clouds around the SMBH, in analogy to the nuclear eclipsing events (e.g., \citealt{2017A&A...607A..28M,2024A&A...684A.101M}), obscure and suppress the intrinsic AGN and TDE emission. This results in a post-flare luminosity that is significantly lower than the pre-flare level. The manuscript is organized as follows. In Sect. \ref{mod}, we present an obscured TDE model. In Sect. \ref{CSS10027}, we apply this model to the light curve of CSS100217. The discussion and conclusions are given in Sect. \ref{diss} and Sect. \ref{con}, respectively. Throughout the 
manuscript, we have adopted the cosmological parameters of $H_{0}$=70 km s$^{-1}$ Mpc$^{-1}$, $\Omega_{m}$=0.3, and 
$\Omega_{\Lambda}$=0.7.

\section{Obscured TDE model}\label{mod}
The observed light curves in AGNs that have experienced a TDE event generally comprise three components: intrinsic AGN activity on sub-parsec scales, the TDE flare on sub-parsec scales, and host-galaxy emission. We propose that the intrinsic emission from the AGN activity and the TDE flare observed along our line of sight can be significantly dimmed due to obscuration by unbound stellar debris produced in the TDE or by moving clouds surrounding the SMBH. Motivated by the fact that the unbound stellar debris is accompanied by a tidal tail or the light variations caused by the observed obscuring clouds are temporally asymmetric \citep{2009MNRAS.400.2070S,2024A&A...684A.101M}, we adopt a time-dependent obscuration model that follows a Weibull distribution. It should be noted that obtaining an exact time-evolution profile of the obscuration is very difficult, as the extinction law produced by unbound stellar debris in TDEs or moving clouds around SMBHs 
is extremely complex. In our model, we assume the extinction law can be approximated by the Galactic extinction law to investigate the effect of obscuration on TDE light curves.

The theoretical TDE model has been discussed in detail in \citet{2013ApJ...767...25G,2014ApJ...783...23G,2018ApJS..236....6G,2019ApJ...872..151M}, and also applied in our recent work in \citet{2024MNRAS.534L..23Z,2025MNRAS.537...84G,2025ApJ...986..174G}. Next, we briefly outline the obscured TDE model to describe the optical light curves.

First, the viscous-delayed accretion rates, $\dot{M}_{a}$, were created based on $dM/dE$ (the fundamental elements in 
the public code of TDEFIT/MOSFIT) and the viscous-delayed effects \citep{2013ApJ...767...25G, 2019ApJ...872..151M}, where $ dM/dE$ is the distribution of debris mass as a function of specific binding energy, $E$, after the star is disrupted. We considered a standard TDE in which a $M_{*}=1{\rm M_\odot}$ main-sequence star is tidally disrupted by a $M_{BH}$= $10^{6}{\rm M}_\odot$ BH. We constructed the standard templates of the time-dependent viscous-delayed accretion rate, $\dot{M}_{a}(T_{v},\beta)$,  
where $\beta=R_t/R_p$ is the impact parameter and $T_{v}$ is the viscous time that arises from both the circularization process and accretion through the disk surrounding the BH, and $R_p$ is the pericenter radius. In addition, we also obtained the time, $t_{a}(T_{v},\beta)$, corresponding to a given $\dot{M}_{a}(T_{v},\beta)$.

Second, for common TDEs, the actual viscous-delayed accretion rate, $\dot{M}$, and the corresponding time, $t$, in the observer frame can be estimated by the following scaling relations \citep{2019ApJ...872..151M}:
\begin{equation}
    \dot{M} \propto M_{\rm BH,6}^{-1/2}\! M_{*}^{2}\! R_{*}^{-3/2}\! \dot{M}_{a}(T_{v},\beta)\;,
\end{equation}
 
\begin{equation}
    t \propto (1+z) M_{\rm BH,6}^{1/2}\! M_{*}^{-1}\! R_{*}^{3/2}\! t_{a}(T_{v},\beta)\;,
\end{equation}
where the central BH mass, $M_{\rm BH,6}$, is in units of ${\rm 10^6M_\odot}$, the stellar mass, $M_{*}$, is in units of ${\rm M_\odot}$,
stellar radius, $R_{*}$, is in units of ${\rm R_{\odot}}$, and $z$ represents the redshift of the TDE host galaxy. Additionally, 
we adopted the known mass-radius relation for main-sequence stars from \cite{1996MNRAS.281..257T}.

Third, following \cite{2019ApJ...872..151M}, we assumed that the radiating region is a simple blackbody photosphere, so the observed flux without obscuration is \begin{equation}
F_{\lambda}(t) = \frac{2\pi hc^2}{\lambda^5} \frac{1}{e^{hc/(k\lambda T_p(t))}-1} \left[\frac{R_p(t)}{D(z)}\right]^2\;,
\end{equation}
with a time-dependent radius of the photosphere of
\begin{equation}
R_p(t) = R_0 \times a_p \left( \frac{L}{L_{\rm Edd}} \right)^{l_p}= R_0 \times \left[ G M_{\rm BH} \left( \frac{t_p}{2\pi} \right)^2 \right]^{1/3} \left( \frac{L}{L_{\rm Edd}} \right)^{l_p}
\end{equation}
and a time-dependent effective temperature of
\begin{equation}
T_p(t) = \left[ {L/}{(4\pi \sigma_{SB} R_p^2}) \right]^{1/4} \;,
\end{equation}
where $c$ is the speed of light, $k$ as the Boltzmann constant, and $D(z)$ is the luminosity distance at redshift $z$. $R_p(t)$ ranges 
from the minimum $R_{\rm isco}$ (innermost stable circular orbit radius) to $a_p$ (maximum semimajor axis) of the accreting mass. $L$ is the 
time-dependent bolometric luminosity given by $L = {\eta\dot{M}(t)c^2}$, $L_{\rm Edd}$ represents the Eddington luminosity given by $L_{\rm Edd} = 1.3\times10^{38}(M_{\rm BH}/M_\odot)\rm \;erg\;s^{-1}$, $\eta$ 
represents the energy transfer efficiency that is less than 0.4 \citep{2014ApJ...783...23G, 2019ApJ...872..151M},  $l_p$ represents the power-law exponent,  $t_p$ is the time of the peak accretion rate, and $\sigma_{\rm SB}$ 
is the Stefan-Boltzmann constant.
However, when line‑of‑sight obscuration is taken into account, the time-dependent  emission spectrum in the rest frame determined by the simple black-body photosphere model is corrected by
\begin{equation}
F_{\lambda}^{\mathrm{obs}}(t) = F_{\lambda}(t) \times b_{\lambda}(t)\;.
\end{equation}
Here $b_{\lambda}(t)\; (<1)$ is a dimensionless parameter that characterizes the time‑dependent extinction (also often called reddening) of the intrinsic spectrum at different wavelengths and is given by \citep{2000ApJ...533..682C} 
\begin{equation}
b_{\lambda}(t)=10^{-0.4k(\lambda)E(B-V)(t)}\;,
\end{equation}
where $E(B-V)(t)$ is the time-dependent color excess and $k(\lambda)$ is the extinction curve. We adopted the Galactic F99 extinction law \citep{1999PASP..111...63F,Gordon2024} to correct for obscuration. This law provides the main general features of extinction from the IR through the UV. 

Finally, to describe the time-dependent evolution of obscuration, we adopted a Weibull distribution as
\begin{equation}
E(B-V)(t) = A \cdot \left( \frac{a}{s} \right) \left( \frac{t - t_0}{s} \right)^{a-1} \exp\left[ - \left( \frac{t - t_0}{s} \right)^a \right],
\end{equation}
where \( A \) is the normalization factor, \( a \) is the shape parameter that determines the sharpness of the obscuration, \( s \) is the scale parameter that determines temporal width of the obscuration, and \( t_0 \) is the time at which extinction begins.
Subsequently, the observed time-dependent apparent magnitudes of the TDE after accounting for obscuration can be determined by convolving the observer-frame model spectrum, $F^{\rm obs}_\lambda(t)$, with the transmission curve of the corresponding photometric filter. In addition, the observed apparent magnitude of the AGN is corrected using the relation
\begin{equation}
\begin{split}
    \rm mag_{obs}(AGN) &= {\rm { mag(AGN)}} + A_V \,, \\
\end{split}
\end{equation}
where $A_V = R_V \times E(B - V)(t)$ with $R_V$ = 3.1 \citep{1999PASP..111...63F}, and $\rm mag(\rm AGN)$ is the intrinsic apparent magnitudes of the AGN. $A_V$ represents the extinction in the $V$ band \citep{1989ApJ...345..245C}. Given that the light curve analyzed in this work was observed in the CSS $V$ band, this relation was adopted to describe the evolution of dust extinction.

In our model, the obscuration effect sets in immediately once the TDE flare is triggered. The free parameters of the model include $M_{\rm BH}$, $M_{*}$, $R_{*}$, $\beta$, $T_{v}$, $\eta$, $R_0$, $l_p$, and the time‑dependent obscuration parameters $A$, $a$, and $s$. Moreover, the parameters $\rm mag(\rm AGN)$ and mag (host) are included to characterize the contributions of the AGN and the host galaxy to the observed variability. 

\section{Application to CSS100217}\label{CSS10027}
\begin{figure*}
\centering\includegraphics[width=15cm,height=10cm]{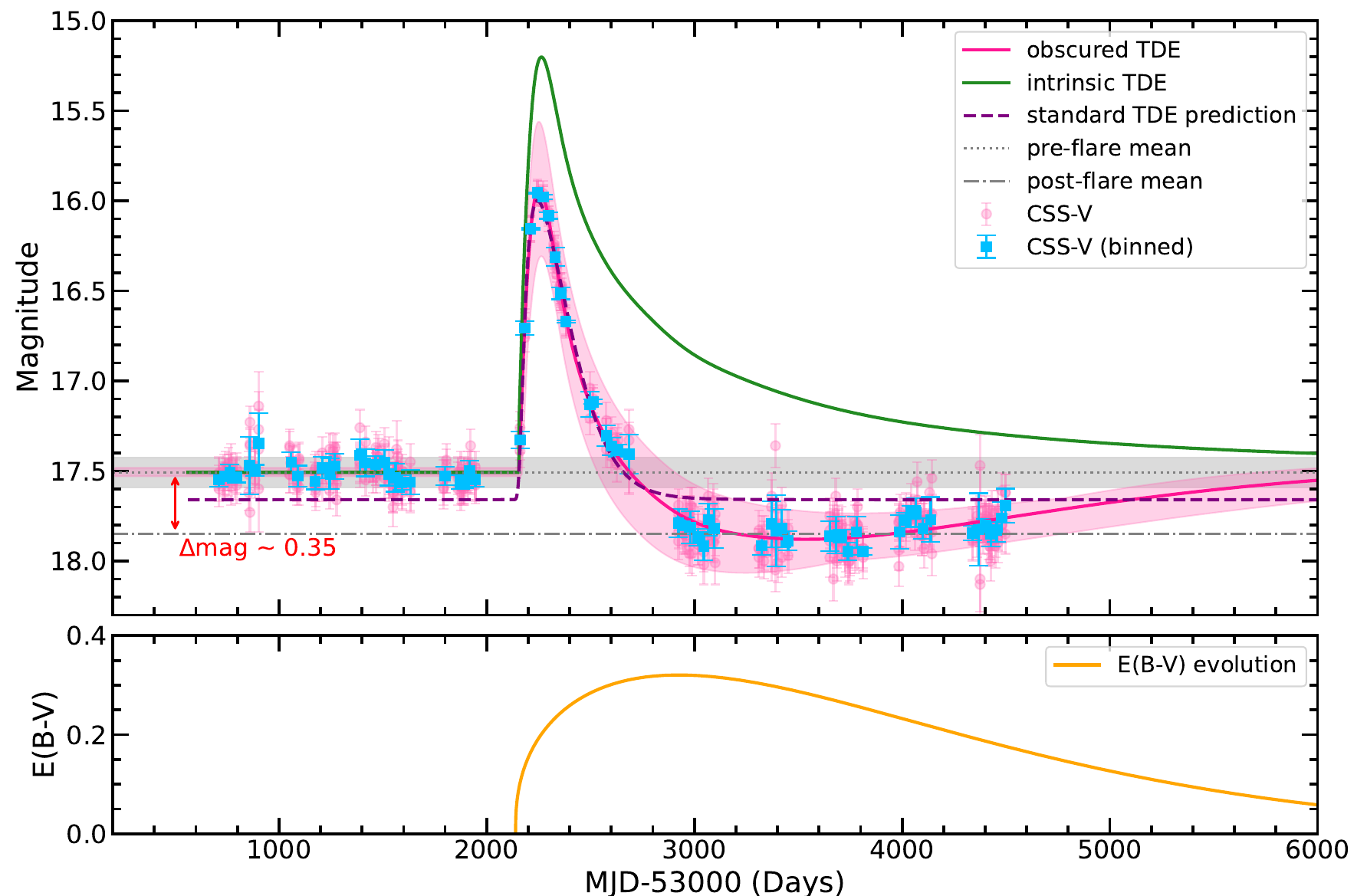}

\caption{Top: Raw single-exposure CSS V-band data are plotted in pink. The binned data are plotted in blue and grouped in sets of 15 consecutive points. The best description by the obscured TDE model is shown as a solid pink line, with the corresponding confidence bands (shaded pink area) determined by uncertainties of the model parameters. For comparison, the solid green line displays the intrinsic TDE light curve recovered by removing the obscured component, while the dashed purple line shows the conventional theoretical TDE model prediction assuming no obscuration. The dashed black line and the gray band show the mean and standard deviation of the pre‑flare magnitudes, respectively, while the dash-dotted gray shows the post‑flare mean magnitude,$\sim$ 0.35 mag fainter (red arrow). Bottom: Time evolution of the color excess E(B–V) determined by the best-fit parameters of the obscured TDE model.}
\label{lc}
\end{figure*}
The long-term photometric V band light curve of CSS100217 was collected from Catalina Sky Survey \citep[CSS;][]{2009ApJ...696..870D} with MJD from 53711 (Nov. 2005) to 57497 (Apr. 2016) and is shown in Fig. \ref{lc}. We applied our obscured TDE model with $t_0=$ 2109 days (MJD-53000) to describe the V band light curve of CSS100217, where $t_0$ represents the beginning of the rising phase of the flare. The model parameters were derived using the maximum likelihood method combined with the Markov chain Monte Carlo (MCMC) technique \citep{2013PASP..125..306F}, leading to the determined model parameters with a polytropic index of $\gamma=4/3$ being:  $\log(M_{\rm BH}/10^6M_\odot)=1.52_{-0.04}^{+0.04}$, $\log(M_{*}/M_\odot)=0.66_{-0.10}^{+0.08}$,  $\log(\beta)=0.19_{-0.03}^{+0.02}$, $\log(T_{v}/\text{years})=-2.07_{-0.09}^{+0.07}$, $\log(\eta)=-0.54_{-0.07}^{+0.12}$,  $\log(R_{0})=-1.31_{-0.04}^{+0.06}$,  $\log(l_{p})=-0.58_{-0.04}^{+0.09}$,  $\log (A/\text{days})=2.92_{-0.05}^{+0.05}$, $a=1.40_{-0.06}^{+0.05}$, $\log(s/\text{days})=3.46_{-0.01}^{+0.01} $, $\rm mag( AGN)=17.75_{-0.02}^{+0.02}$, and $\rm mag (host)=19.23_{-0.06}^{+0.05}$. The determined posterior distributions of the model parameters are shown in Fig. \ref{MCMC}. The best-fitting results and the 1$\sigma$ confidence bands derived from the  uncertainties of the model parameters are shown in the top panel of Fig. ~\ref{lc}. The reduced $\chi2 / dof$ of our fit is $\sim$ 0.8. The V‑band light curve of CSS100217 can be described well by an obscured TDE, in which a main-sequence star with a mass of $4.6_{-0.9}^{+0.9}{\rm M_\odot}$ is tidally disrupted by the SMBH with a mass of $3.3_{-0.3}^{+0.3}\times10^7{\rm M_\odot}$. The total energy of the event derived from our fit is $7.23\times10^{53}$ ergs. About 1.38 ${\rm M_\odot}$ of debris is accreted by the central SMBH. Our analysis shows that the host galaxy  is about 1.5 magnitudes fainter than the AGN, indicating that the pre-flare luminosity is dominated by the AGN.

Based on the best‑fitting parameters of the obscured TDE model, we also present the intrinsic TDE flare without obscuration (solid green line in the top panel of Fig.\ref{lc}). We find that the obscuration causes the intrinsic TDE flare to decline rapidly from its peak to a level below the pre-flare, with the strongest obscuration occurring at MJD-53000$\sim$ 2900 days, as is shown by the solid orange line in the bottom panel of Fig. \ref{lc}. Our model further predicts that the low‑luminosity state will return to pre‑flare levels at  MJD-53000 $\sim$ 6000 days.

\section{Discussion}\label{diss}
We present the SDSS-provided spectrum of SDSS J102911.94+404220.8 in Fig. \ref{spe}.1. Spectral analysis indicates that the radiation of this object almost entirely originates from the core components rather than the host galaxy. This is consistent with the results of the light curve analysis. In addition, the BH mass derived from our fit is consistent with the value of $\sim 10^7{\rm M_\odot}$ reported by \citet{2022MNRAS.516..529C} based on the continuum luminosity and the flux and width of the $\rm {H\beta}$ broad emission line.

The unbound stellar debris during a TDE would generally be concentrated in a cone or “fan” close to the orbital plane (e.g., \citealt{1988Natur.333..523R,2009MNRAS.400.2070S,2019GReGr..51...30S}). \citet{2009MNRAS.400.2070S} showed that the unbound debris is distributed along an arc, with dispersion in specific energy producing a spread in both radius and azimuthal angle within the orbital plane (see Fig.1 in \citealt{2009MNRAS.400.2070S}). The most energetic unbound stellar debris is ejected at speeds of $v_{\rm max}=(3R_*/R_{\rm p})^{1/2}v_{\rm p}$, where $v_{\rm p}=(2GM_{BH}/{R_{\rm p}})^{1/2}$ is the orbital velocity at pericenter.  
Figure \ref{rad} shows that the radius of the unbound stellar debris always exceeds the TDE photospheric radius.
At maximum obscuration (about 1000 days after disruption), the unbound stellar debris lies at $R \sim 0.1$ pc from the central BH. About $M=3.14\;{\rm M_\odot}$ of the material is ejected as unbound stellar debris. This unbound stellar debris subtends a solid angle, $\Delta\Omega\sim48^{1/2}(R_*/R_{\rm p})^{3/2}\sim0.002$ sr \citep{2009MNRAS.400.2070S}. The number density of particles in the cone of unbound debris can be estimated as $n\sim (M/m_{p})/(R^2\Delta R\Delta\Omega/3)\sim2\times10^9 \;\rm cm^{-3}$, where $\Delta R\sim R (3R_*/R_{p})^{1/2}$ is the radial dispersion of the debris at a fixed azimuthal angle \citep{2009MNRAS.400.2070S}. The density of the radial column can be estimated as $N\sim n\cdot\Delta R\sim5.6\times10^{26}\;\rm cm^{-2}$. As the unbound debris expands, it cools very quickly. After a few months, the gas would be completely neutral and may condense into dust if not for the ionizing radiation of the disk.  Therefore, if the unbound stellar debris lies along our line of sight, it can significantly dim the intrinsic AGN activity and the TDE flare. Furthermore, \cite{2024A&A...684A.101M} observed that moving clouds with a column density near $10^{22\sim23} \; \rm cm^{-3}$ around SMBHs induce variations of $\sim$ 0.2 magnitudes in the V band and $\sim$ 0.4 magnitudes in the B band of the accretion disk.

We have collected the following optical data of CSS100217 from the \textit{Zwicky} Transient Facility \citep[ZTF;][]{2019PASP..131a8002B} and the Asteroid Terrestrial-impact Last Alert System (ATLAS) forced photometry \citep{2021TNSAN...7....1S}, as is shown in Fig. \ref{LC2}.1. Comparing phometric results of different facilities that use different filters is nontrivial. Nevertheless, the follow-up observations show an overall stable trend with insignificant variability. The model predicts a modest rise before reaching the pre-flare level. This may be because the actual $E(B-V)$ becomes fainter more slowly than is inferred from the fit.

\section{Conclusions}\label{con}
CSS100217 has been reported in the literature as a transient for which a TDE interpretation has been favored. Notably, the brightness of CSS100217 is significantly lower post-flare than pre-flare. In this work, we propose an obscured TDE scenario as the origin of the observed dimming, presenting a viable explanation for the atypical light curve of CSS100217. Our analysis shows that the AGN dominates the observed optical level in the pre-flare of CSS100217. The BH mass derived from our fit is consistent with the value of $\sim 10^7{\rm M_\odot}$ reported by \citet{2022MNRAS.516..529C}.  According
to the obscured TDE scenario, the post-flare brightness is predicted to return to its pre-flare level $\sim$ 4000 days after the flare peak. The results presented in this manuscript could be conveniently applied to describe a unique class of TDEs whose post-flare luminosities are lower than their pre-flare levels.

\begin{acknowledgements}
This work is supported by the National Natural Science 
Foundation of China (grants NSFC-12173020, 12373014 and 12133003). Gu gratefully thank the kind financial support from the 
Innovation Project of Guangxi Graduate Education. The paper has made use of the code of MOSFIT (Modular Open Source Fitter for Transients) 
\url{https://mosfit.readthedocs.io/} which is a Python 2.7/3.x package for fitting, sharing, and estimating the parameters 
of transients via user-contributed transient models. The paper has made use of the MCMC code 
\url{https://emcee.readthedocs.io/en/stable/index.html}.
\end{acknowledgements}

\bibliographystyle{aa} 
\bibliography{references} 

\onecolumn
\begin{appendix}
\section{Posterior distributions of obscured TDE model parameters}

\renewcommand{\thefigure}{A.1}
Figure \ref{MCMC} shows the posterior distributions of the TDE model parameters obtained from the MCMC technique. 
\begin{figure*}[h!]
\centering
     \includegraphics[width=18cm, height=20cm]{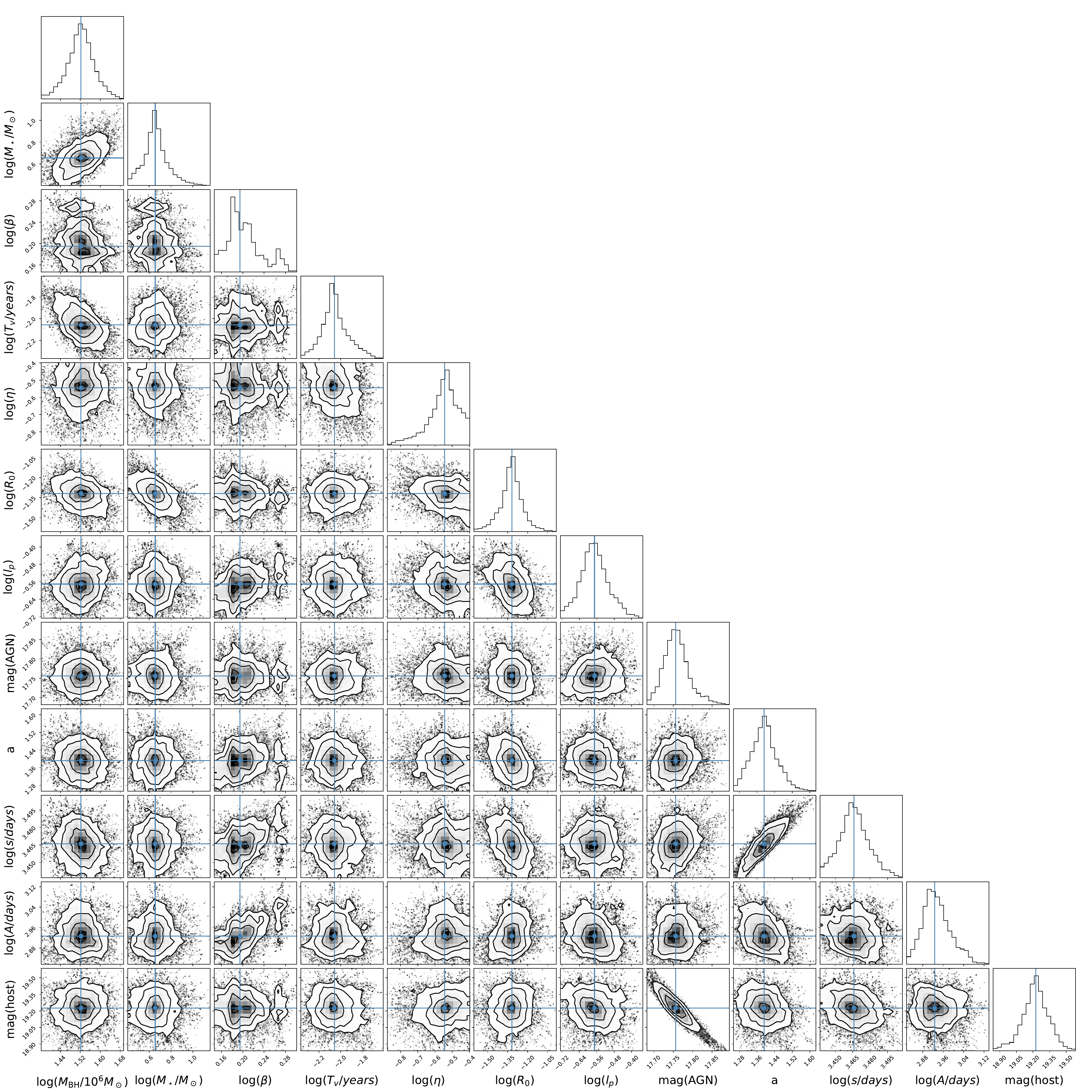}
  \caption{Posterior distributions of the TDE with
   obscured model parameters derived from the MCMC technique. In each panel, the three circles from outer to inner represent $3\sigma$, $2\sigma$, and $1\sigma$ confidence levels and the blue dot in the center of each contour marks the position of the best-fit parameter.}\label{MCMC}
\end{figure*}
\clearpage

\twocolumn
\section{The spectrum of SDSS J102911.94+404220.8}

Figure \ref{spe}.1 presents the SDSS-provided spectrum of SDSS J102911.94+404220.8. As shown in the right panel of Fig. \ref{spe}.1, we cannot find obvious absorption line features in the common distribution range 
(3900$\sim$4100 \AA), due to that the Ca~\textsc{ii} H \& K  features are hidden under the [Ne~\textsc{iii}] emission line. 
However, the continuum slope and emission lines clearly indicate the presence of an AGN, presenting evidence that the spectrum is AGN-dominated.
This indicates that the radiation of this object almost entirely originates from the core components rather than the host galaxy.

\begin{figure}[!htbp]\label{spe}
\renewcommand{\thefigure}{B.1}
\includegraphics[width=0.43\paperwidth]{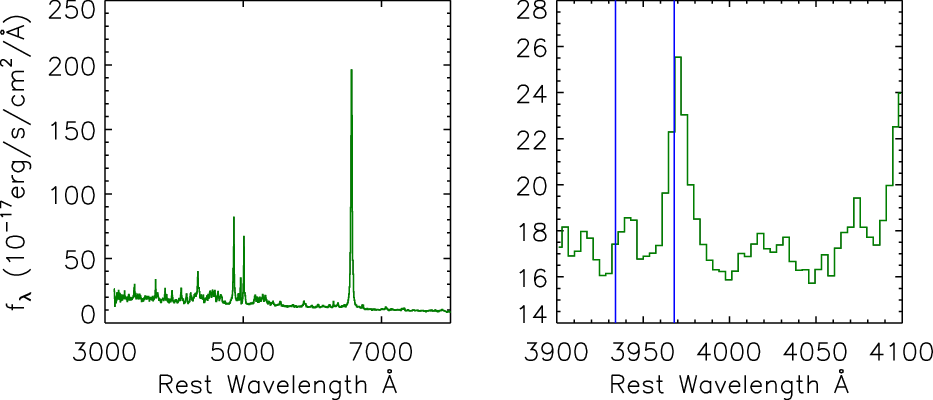}
 \centering
 \caption{Left: Rest frame spectrum of CSS100217. 
Right: Spectrum in the range 3900$\sim$4100 \AA. 
In both panels, the dark green components indicate the spectrum. 
In the right panel, the vertical blue lines mark the center wavelengths of Ca~\textsc{ii} H \& K absorption lines. 
}
\end{figure}

\section{Optical observations from ATLAS and ZTF}
Long-term light curves were collected from ZTF (MJD 58206–60092; March 2018–May 2023) and ATLAS (MJD 57316–60823; October 2015–May 2025), as shown in Fig. \ref{LC2}.1.

\renewcommand{\thefigure}{C.1}
\begin{figure}[!htbp] \label{LC2}
\includegraphics[width=0.42\paperwidth]{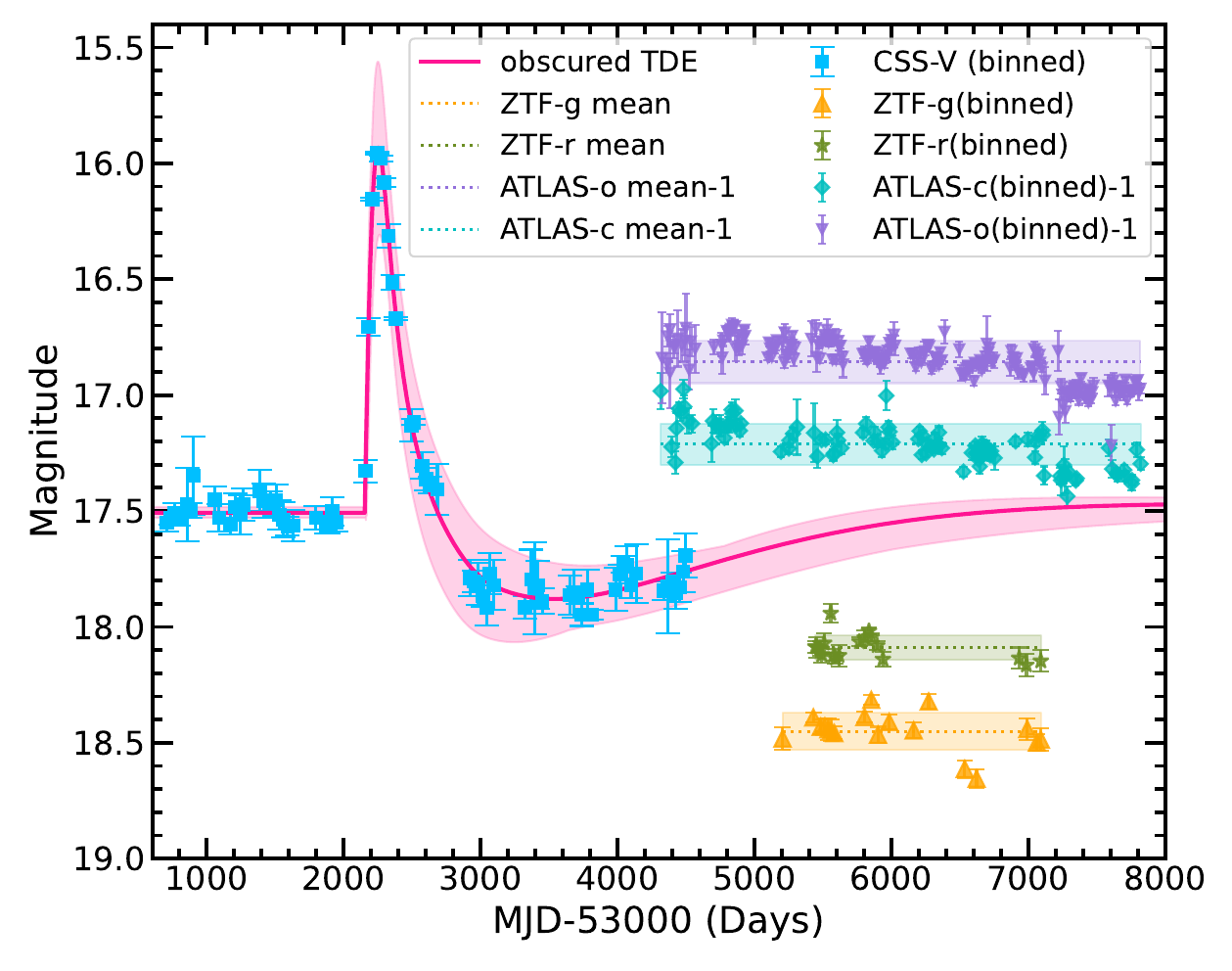}
 \centering
\caption{Same as Fig. \ref{lc} but showing follow-up observations from other projects, as indicated in the legend at the top right. All data points are shown after 15-day binning. The dashed line and corresponding shaded region of the same color indicate the mean value and standard deviations of the light curve. For clarity, the ATLAS data points, as well as the corresponding mean and standard deviation, have been offset by $-1$~mag.
 }
\end{figure}

\section{Comparison of the photospheric radius with the unbound stellar debris radius}
In our model, we assume that obscuration occurs at the onset of the flare. The flare is proportional to the accretion rate of the BH ($\dot{M}_{a}$). In a TDE, 
due to the time delays introduced by the circularization process and by accretion through the disk surrounding the BH, 
\begin{equation}
    \dot{M}_{a}(t)=\dot{M}_{fb}(t)-M_a(t)/T_v \;,
\end{equation} where $T_v$ is the viscous timescale, $\dot{M}_{fb}(t)$ is the debris fallback rate, $M_a$ is the mass of debris that remains suspended outside of the BH's horizon for roughly a viscous time \citep{2019ApJ...872..151M}.
The viscous timescale $T_v$ derived from our model is $\sim 3$ days. Thus, at the onset of obscuration, the unbound stellar debris lies at a distance of $v_{\rm max}T_v\sim0.0002$ pc from the central BH, which is larger than the photospheric radius of the TDE at that time, as shown in Fig. \ref{rad}.1. Subsequently, as time evolves, the radius of the unbound stellar debris always exceeds the TDE photospheric radius.
\renewcommand{\thefigure}{D.1}
\begin{figure}[!htbp]\label{rad}
\includegraphics[width=0.4\paperwidth]{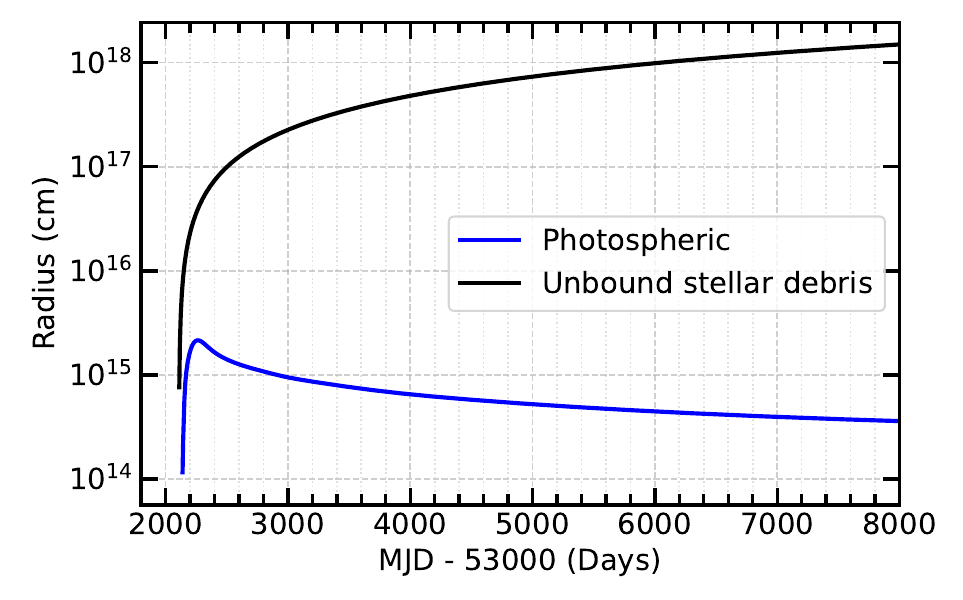}
 \centering
\caption{Time evolution of the photospheric radius of the TDE and the radius of the unbound stellar debris.}
\end{figure}

\end{appendix}
\end{document}